\documentstyle[aps,pre,twocolumn,amsfonts,floats,graphicx]{revtex}

\begin{document}

\draft

\preprint{Submitted to Physical Review E }

%%% This ``wideabs'' trick has to be removed before sending the
%%% compuscript to the APS offices!!!
%%% Note that \footnote and \thanks will disappear.
%\wideabs{
  
\title{Discrete breathers in dissipative lattices}
  
\author{J. L. Mar\'{\i}n$^{(a),(c)}$, F.~Falo$^{(a),(c)}$,
  P.~J.~Mart\'{\i}nez$^{(b),(c)}$ and L.~M.~Flor\'{\i}a$^{(a),(c)}$}
  
\address{$^{(a)}$ Dept.\ de F\'{\i}sica de la Materia Condensada,
  Universidad de Zaragoza, 50009~Zaragoza, Spain}
  
\address{$^{(b)}$ Dept.\ de F\'{\i}sica Aplicada, Universidad de
  Zaragoza, 50009~Zaragoza, Spain}
  
\address{$^{(c)}$ Dpto.\ de Teor\'{\i}a y Simulaci\'on de Sistemas
  Complejos\\
  Instituto de Ciencia de Materiales de Arag\'on\\
  C.S.I.C. - Universidad de Zaragoza, 50009~Zaragoza, Spain}
  
\date{\today}
  
\maketitle

\begin{abstract}
  We study the properties of discrete breathers, also known as
  intrinsic localized modes, in the one-dimensional Frenkel-Kontorova
  lattice of oscillators subject to damping and external force.  The
  system is studied in the whole range of values of the coupling
  parameter, from $C=0$ (uncoupled limit) up to values close to the
  continuum limit (forced and damped sine-Gordon model).  As this
  parameter is varied, the existence of different bifurcations is
  investigated numerically.  Using Floquet spectral analysis, we give
  a complete characterization of the most relevant bifurcations, and
  we find (spatial) symmetry-breaking bifurcations which are linked to
  breather mobility, just as it was found in Hamiltonian systems by
  other authors.  In this way moving breathers are shown to exist even
  at remarkably high levels of discreteness.  We study mobile
  breathers and characterize them in terms of the phonon radiation
  they emit, which explains successfully the way in which they
  interact.  For instance, it is possible to form ``bound states'' of
  moving breathers, through the interaction of their phonon tails.
  Over all, both stationary and moving breathers are found to be
  generic localized states over large values of $C$, and they are
  shown to be robust against low temperature fluctuations.
\end{abstract}


%}%%% End of the ``wideabs'' environment. Remember to remove.

\section{INTRODUCTION}

The phenomenon of (non-topological) localization in discrete nonlinear
lattices ({\em i.e.\/} intrinsic localized modes or discrete
breathers) has received a great deal of attention from both
theoretical and (as of lately) experimental research.  Indeed, recent
observations~\cite{TMO00,BAUFZ00,BAU00} of discrete roto-breathers in
Josephson-junction ladder circuits have placed the subject on a firm
experimental footing (see also~\cite{SES99}).  Most of the theoretical
and computational work on discrete breathers has dealt with
Hamiltonian systems, of fundamental interest in Physics. For a review,
see Refs.~\cite{FW98,Aub97}.  Comparatively, the easier case of
dissipative breathers has received much less attention, though the
experimental systems that we have just mentioned belong to this class.

Mathematical proofs of existence of discrete breathers in rather
general dissipative networks of oscillators~\cite{SM97,Aub97} appeared
soon after those of Hamiltonian networks~\cite{MA94}.  While in the
later case, a condition of non-resonance of the localized oscillation
with the band of extended normal modes of the lattice has to be
satisfied, that is not an issue in the forced-damped case, and the
dissipative breather possesses the character of attractor for initial
conditions in the corresponding basin of attraction.  As archetypical
example of Klein-Gordon lattices of oscillators, we consider the
standard Frenkel-Kontorova model with commensurability one ({\em
  i.e.\/} average interparticle distance equal to the period of the
sinusoidal substrate potential).  In section~\ref{sec:model} we
discuss the numerical procedures used to obtain accurate breather
solutions, which are based on the continuation from the uncoupled
limit of the model.

In section~\ref{sec:stability} we explain some general features of the
linear stability (Floquet) analysis of forced-damped periodic discrete
breathers.  For the sake of readability, this section is intended to
be self-contained, to some extent.  After deriving some
straightforward properties of the Floquet multipliers, we obtain some
formal conditions for the non-appearance of extended instabilities of
the uniformly oscillating background, along with the tail analysis
valid for not-too-large forcing.  The good fitting of our numerical
data to the results of this section ensures its validity for the
parameters used in the numerical work.

Section~\ref{sec:bifurcations} reports on our numerical findings,
concerning pinned discrete breathers, which are summarized in the
phase diagram against the coupling parameter.  Pitchfork and
Andronov-Hopf bifurcations separating different periodic and
quasiperiodic breathers appear as generic features of this phase
diagram.  At very high values of the coupling parameter, when the
width of the discrete breather is much larger than the period of the
substrate potential, a Goldstone mode in the Floquet spectrum signals
the approach to the continuum limit.

In section~\ref{sec:mobile} we study the mobility of discrete
breathers, a subject which is yet poorly understood.  After discussing
the procedures used to obtain and continue mobile breathers we explain
successfully, with the aid of simple physical arguments, the numerical
power spectra in the tails.  Then we study collisions between discrete
breathers.  We find all possible scenarios, ranging from ``elastic''
to completely ``inelastic'' collisions; this latter case includes both
breather annihilation and, more interestingly, the formation of
``breather molecules'' which can be either pinned or mobile.  Finally,
in section~\ref{sec:conclusions}, we summarize the main conclusions of
our work.

\section{MODEL AND BREATHER GENERATION}
\label{sec:model}

The equations of motion of the Frenkel-Kontorova chain subject to
damping and an (spatially uniform) external driving force are, in
dimensionless form,

\begin{eqnarray}  
  \ddot{u}_j + \alpha\dot{u}_j + &&\frac{1}{2\pi}\sin(2\pi u_j) =
  \nonumber\\
  &&C\left(u_{j+1}-2u_j+u_{j-1}\right) + F_{ac}\sin(\omega_b t)
  \label{eq:mot}
\end{eqnarray}

In order to generate a discrete breather configuration we start in the
anti-integrable (uncoupled) limit $C=0$, using two different amplitude
attractors of the single pendulum equation of motion.  That is, we
first consider the dynamics of a single forced and damped pendulum,
and try to find a region of parameters where there is at least two
different attractors coexisting.  Note that, generically, all
oscillators have at least two attractors for sufficiently low values
of the damping $\alpha$ and the force $F_{ac}$, if the frequency of
the force $\omega_b$ is not wildly different from the typical
frequencies of the autonomous oscillator.

Therefore we initially choose values for $\alpha$, $F_{ac}$, and
$\omega_b$, and keep them fixed while we vary $C$.  Then, for
instance, we fix one of the oscillators to the high amplitude solution
and all the others to the low one.  Using as initial condition this
anti-integrable configuration, we turn on adiabatically the coupling
parameter $C$.  Following Sepulchre and MacKay's work on dissipative
breathers~\cite{SM97}, the initial solution can be continued for
$C\neq0$, at least until a bifurcation is reached.  That paper shows
how, in contrast to Hamiltonian systems, forced and damped systems
have it easier to comply with the conditions of the continuation
theorem, since there is no problem of resonance with phonons (we have
attractors), and the relative phases of the oscillators are locked by
the external force.

Moreover, if the variation in $C$ is small enough, the discrete
breather remains an attractor of the dynamics (since one expects the
basins of attraction to evolve continuously with $C$ as well).  This
makes the numerical continuation greatly simpler: it is possible to
just vary adiabatically the coupling $C$ as we integrate the equations
of motion~(\ref{eq:mot}), and the dissipative dynamics drives the
system to the stable attractor.  Contrast this with the expensive
root-finding methods one needs to use for breather continuation in
Hamiltonian problems~\cite{MA96}.

In addition, we performed a linear stability analysis of the periodic
solutions (Floquet-Bloch analysis, see below) in order to investigate
the nature of possible bifurcations.  In some cases we have added to
the initial conditions a small random noise (typically of order
$10^{-5}$) to test for robustness.  We have taken special care dealing
with finite size effects.  For low values of $C$ ($C<0.6$) small
lattice sizes can be used (say $N=40$).  However, once the breather is
dressed by a phonon tail (see below), we have needed to increase the
lattice size up to $N=900$ in order to avoid finite size effects.  We
think this is an important point to check since experiments in real
dissipative systems are done in small lattices~\cite{TMO00,BAUFZ00}.
Numerical integration of equations of motion has been done using a
fourth-order Runge-Kutta scheme.  Most of the simulations in this
paper have been done with the following parameters: $\alpha=0.02$,
$\omega_b=0.2\pi$, and $F_{ac} = 0.02$, although we have sometimes
changed them to confirm the general validity of our results.

\section{LINEAR STABILITY ANALYSIS}
\label{sec:stability}

\subsection{Floquet multipliers}

Let us consider a small perturbation, $\{v_j(t)\}$, of the discrete
breather $\{u_j(t)\}$ solution, $v_j= u_j+\epsilon_j$. After direct
substitution in the equations of motion and discarding terms which are
nonlinear in $\epsilon_j$, one finds

\begin{equation}
  \ddot{\epsilon}_j + \alpha\dot{\epsilon}_j +
  \cos(2\pi u_j(t))\epsilon_j = C \left(
    \epsilon_{j+1}-2\epsilon_j+\epsilon_{j-1} \right)
  \label{eq:linmot}
\end{equation}

These form a system of coupled linear differential equations with time
periodic coefficients, for $u_j(t)$ is a periodic function of time.
For a system of size $N$, the integration of the linearized
equations~(\ref{eq:linmot}) over a period $t_b= 2\pi/\omega_b$ of each
of the $2N$ vectors $\{\epsilon_j(0),\dot{\epsilon}_j(0)\}$ forming
some basis of the tangent space defines the $2N\times 2N$ Floquet (or
monodromy) matrix $\cal F$
\begin{equation}
  \left(
    \begin{array}{c}
      \epsilon_j(t_b) \\
      \dot{\epsilon}_j(t_b)
    \end{array}
  \right)=
  \cal{F} \left(
    \begin{array}{c}
      \epsilon_j(0) \\
      \dot{\epsilon}_j(0) 
    \end{array}
  \right)
  \label{eq:floquet}
\end{equation}
that relates the small perturbations at $t=t_b$ to those at $t=0$; in
other words, $\cal F$ is the matrix associated to the $t_b$-map of
(\ref{eq:linmot}).

The linear stability of the breather solution $\{u_j(t)\}$ requires
that all the eigenvalues of the Floquet matrix (called also {\em
  Floquet multipliers}) are inside the unit circle. Since $\cal F$ is
real, if $\mu$ is an eigenvalue of $\cal F$, its complex conjugate
$\overline{\mu}$ is also an eigenvalue of $\cal F$.  But the Floquet
spectrum has more structure, since one can transform the linear system
of Eq.~(\ref{eq:linmot}) into a Hamiltonian one (see
Ref.~\cite{WvdZSO96}).  By transforming the $\epsilon_j$ variables
according to
\begin{equation}
  \epsilon_j(t)= e^{-\alpha t/2} \eta_j(t),
  \label{eq:transformation}
\end{equation}
this yields
\begin{eqnarray}
  \ddot{\eta}_j -
    \Bigl( \alpha^{2}/4-\cos\bigl(2\pi u_j(t) && \bigr) \Bigr)  \eta_j =
  \nonumber\\
  && C ( \eta_{j+1} - 2\eta_j + \eta_{j-1} )
  \label{eq:transf}
\end{eqnarray}
These are the equations of motion of a (non autonomous) Hamiltonian
system of oscillators, for which the eigenvalues of the (symplectic)
$t_b$-map must come in pairs such that their product is unity.
Together with the fact that the map is real, one has these well
known~\cite{Arn89} three possible cases: {\em(i)} pairs of eigenvalues
lying on the unit circle, with $\lambda_1=\overline{\lambda}_2$;
{\em(ii)} pairs lying on the real axis, with $\lambda_1=1/\lambda_2$;
{\em(iii)} 4-tuples of eigenvalues with
$\lambda_1=\overline{\lambda}_3$, $\lambda_2=\overline{\lambda}_4$,
$\lambda_1=1/\overline{\lambda}_4$.

Since the transformation (\ref{eq:transformation}) scales the
eigenvalues by a factor $\exp[-\alpha t_b/2]$, the Floquet multipliers
of (\ref{eq:linmot}) must either lie on a circle of radius
$\exp(-\alpha t_b/2)$, or on the real axis such that
$\mu_1\mu_2=\exp(-\alpha t_b)$, or come as 4-tuples such that
$\mu_1=\overline{\mu}_3$, $\mu_2=\overline{\mu}_4$,
$\mu_1=\exp(-\alpha t_b)/\overline{\mu}_4$.

An important difference with the Hamiltonian case, where a ``phase''
and the ``growth'' modes~\cite{Aub97} are always associated to the
double eigenvalue $+1$ in the Floquet matrix of the discrete breather,
is that these modes do not exist for the forced-damped case.  The
reason for that is that both the breather frequency and the time
origin are fixed by the external force, so that the associated
degeneracies are removed.

\subsection{Extended instabilities}

In the limit of an infinite system ($N\rightarrow \infty$), the
spectrum of $\cal F$ consists of a continuous part associated with
spatially extended eigenvectors and a discrete part associated with
spatially localized eigenvectors. The continuous part of the spectrum
of $\cal F$ is the continuous spectrum of the linearized problem
around the homogeneous solution ({\em i.e.}, without breather)
$\{u_j(t)\} = \{u_{\infty}(t)\}$. As pointed out by Mar\'{\i}n and
Aubry \cite{MA98}, using the fact that the limit (in the appropriate
sense) of the sequence of spatial translations of the Floquet matrix
$\cal F$ of the system with breather is the Floquet matrix ${\cal
  F}_0$ of the system without the breather, one proves easily that the
spectrum of ${\cal F}_0$ is included in the spectrum of $\cal F$.
Reciprocally, the limit of the sequence of spatial translations of an
extended eigenvector of $\cal F$ can be seen to belong to the spectrum
of ${\cal F}_0$.

First, we are going to consider the spectrum of ${\cal F}_0$, so we
now will pay attention to the linearized equation of motion
(\ref{eq:linmot}) around the homogeneous solution of (\ref{eq:mot}),
$\{u_j(t)\} = \{u_{\infty}(t)\}$ and denote simply $f(t) = \cos(2\pi
u_{\infty}(t))$.  Under the usual periodic boundary conditions, we
look for solutions of the linear problem with the plane-wave form
\begin{equation}
  \eta_j(t) = e^{iqj} \chi^{q}(t)
  \label{eq:chi}
\end{equation}

In other words, $\chi^{q}(t)$ is the (spatial) Fourier coefficient of
$\eta_j(t)$.  Inserting (\ref{eq:chi}) into the equations
(\ref{eq:linmot}), and denoting by $E(q) = 4C\sin^{2}(q/2) -
\alpha^{2}/4$, we have, for each value of $q$, the equation
\begin{equation}
  \ddot{\chi}^{q}(t) + \bigl( E(q) +f(t) \bigr) \chi^{q}(t) = 0
  \label{eq:Hill}
\end{equation}

This is a Hill equation. For each solution $\chi^{q}(t)$ of the single
Hill equation (\ref{eq:Hill}) we have a solution of the form
(\ref{eq:chi}) for the equations (\ref{eq:transf}), and thus, a
solution
\begin{equation}
  \epsilon_j(t) = e^{iqj} e^{-\alpha t/2} \chi^{q}(t) 
  \label{eq:homogeneous}
\end{equation}
for the linearized problem. The Hill equation (\ref{eq:Hill}) has a
general solution which can be expressed in terms of its {\em normal
  solutions}, which have the property
\begin{equation}
  \chi^{q}(t+2\pi/\omega_b) = \lambda_{q} \chi^{q}(t),
  \label{eq:characteristic}
\end{equation}
where $\lambda_{q}$ is called the characteristic number of the
equation.  The complex number $\rho_{q}$ defined as $\lambda_{q}=
\exp(2\pi\rho_{q}/\omega_b)$ is the called characteristic exponent
(its imaginary part being defined up to an additive multiple of
$\omega_b$).  In the generic case in which equation (\ref{eq:Hill})
has two different characteristic numbers $\lambda_{q}^{+},
\lambda_{q}^{-}$, their product is equal to unity, $\lambda_{q}^{+}
\lambda_{q}^{-}= 1$, and the general solution has the form
\begin{equation}
  \chi^{q}(t) = c_{+} e^{\rho_{q}^{+} t} \psi_{q}^{+}(t) + 
    c_{-} e^{\rho_{q}^{-} t} \psi_{q}^{-}(t)
  \label{eq:general}
\end{equation}
where $c_{+},c_{-}$ are constants and $\psi_q^{+},\psi_q^{-}$ are time
periodic functions with period $2\pi/\omega_b$.  Consequently,
$\chi^q(t)$ is bounded by $K\exp(\rho^{\text{max}}_q t)$, with $K$
some constant, and $\rho^{\text{max}}_q =
\max\{\rho_{q}^{+},\rho_{q}^{-}\}$.  Thus, from equation
(\ref{eq:homogeneous}), we conclude that the stability of the
homogeneous solution $\{u_{\infty}(t)\}$ is assured in the parameter
region in which
\begin{equation}
  \rho_{\text{sup}}= \sup_q \rho_q^{\text{max}}  <  \alpha/2
\end{equation}

The determination of this region in parameter space can only be made
by numerical means. For the range of parameters that we have used in
our study of damped-forced breathers, the function $f(t)$ is a low
amplitude oscillation around the value $1$, and, as expected from the
well-known results on weakly time dependent Hill equations, we have
not observed instabilities by extended modes.

In the next section, we will follow the continuation of the breather
solution from the uncoupled limit, for increasing coupling and
numerically compute the eigenvalues of the Floquet matrix ${\cal F}$.
This will allow the characterization of the different bifurcations
that the breather experiences when the coupling parameter increases.

\subsection{Tail analysis}
\label{subsec:tail}

To proceed a bit further, we will assume from now on in this section
that $u_{\infty}(t)$ is an oscillation of very low amplitude, so that
for $|j|\gg 1$ the coefficient $\cos(2\pi u_j(t))$ in equations
(\ref{eq:linmot}) is essentially unity, if one discards terms less
than or equal to $u^{2}_{\infty}(t)$.  Then we are left with the
standard problem of a linear chain with damping, which we can solve
exactly (a similar analysis to the one below appears in
Ref.~\cite{FS99}).

Let us consider a semi-infinite chain with the boundary condition at
the beginning given by $\epsilon_0(t) = \exp(-i\omega t)$, and look
for solutions of (\ref{eq:linmot}) of the form
\begin{equation}
  \epsilon_j(t) = e^{(-\xi+iq)j} e^{-i\omega t} 
  \label{eq:ansatz}
\end{equation}

Inserting (\ref{eq:ansatz}) into equations (\ref{eq:linmot}) one
obtains for the real and imaginary part, respectively
\begin{mathletters}
  \label{eq:dispersion} 
  \begin{eqnarray}
    \label{eq:dispersiona}
    \cosh\xi \cos q  & = &  1+ \frac{1}{2C}(1-\omega^{2}) \\
    \sinh\xi \sin q  & = &  \frac{\alpha\omega}{2C}
    \label{eq:dispersionb}
  \end{eqnarray}
\end{mathletters}

First, we will analyze the situation in which $\alpha = 0$, and see
how one recovers the well-known results for Hamiltonian discrete
breathers~\cite{Mar97,CAF98}.  For $\xi =0$, one has the familiar {\em
  normal mode} solutions, where the frequencies are given by the
dispersion relation
\begin{equation}
  \omega^{2} = 1 + 4C\sin^{2}(q/2)
  \label{eq:disprel}
\end{equation}
and $ q$ ($-\pi < q < \pi$) is the wave-vector of the normal mode. It
is customary to denote loosely the normal modes as ``phonons'', and
the interval of values of $\omega$ defined by (\ref{eq:disprel}) as
``phonon band''.  For $\xi \neq 0$ one has exponentially decaying
solutions
\begin{mathletters}
  \begin{eqnarray}
    \epsilon_j(t) & = & e^{-\xi j} e^{-i\omega t} \\
    \epsilon_j(t) & = & (-1)^{j} e^{-\xi j} e^{-i\omega t}
  \end{eqnarray}
\end{mathletters}
where the inverse decay length $\xi$ and the frequency $\omega$ are
related, respectively, through:
\begin{mathletters}
  \label{eq:decay}
  \begin{eqnarray}
    \label{eq:decaya}
    \omega^{2} & = & 1 - 4C\sinh^{2}(\xi/2) \\
    \omega^{2} & = & 1 + 4C\cosh^{2}(\xi/2) 
    \label{eq:decayb}
  \end{eqnarray}
\end{mathletters}

Note that the values of $\omega$ in (\ref{eq:decay}) are,
respectively, below and above the phonon band, so we observe how the
Hamiltonian linear lattice damps out any solution with a frequency
component outside the phonon band (\ref{eq:disprel}), while the normal
modes are extended ($\xi=0$). As a consequence, a Hamiltonian breather
needs to have all breather harmonics $n\omega_b$ out of the phonon
band, and then they decay exponentially with the characteristic length
$\xi^{-1}(n\omega_b)$.  Thus the size of the Hamiltonian breather is
given by $\xi^{-1}_b= \sup_n \xi^{-1}(n\omega_b)$.

When $\alpha \neq 0$, we have that $\xi(\omega) \neq 0$. Thus, any
solution decays exponentially. For very low values of the damping, and
frequencies well inside the (Hamiltonian) phonon band, the decay
length $\xi^{-1}$ is very large, so that $\sinh\xi\simeq\xi$ in
equation (\ref{eq:dispersionb}), and thus it can be approximated by
\begin{equation}
  \xi^{-1} \simeq \frac{2C\sin q}{\alpha\omega} = \frac{2v_g}{\alpha}
  \label{eq:approx}
\end{equation}
where $v_g = (d\omega/dq)$ is the ``group velocity'' of the
corresponding normal mode, obtained from the dispersion relation
(\ref{eq:disprel}). This approximation admits a simple physical
interpretation in terms of the competition between the damping and the
velocity $v_g$ at which the wave generated (at the beginning of the
semi-infinite chain) by the sustained perturbation propagates: The
amplitude of the excited phonon decays in time as $\exp(-\alpha t/2)$,
so that the time after which the amplitude has decayed by a factor of
($1/e$) is $2/\alpha$, and thus the distance traveled by the phonon is
$2v_g/\alpha$.

\begin{figure}
  \begin{center}
    \includegraphics[width=0.45\textwidth]{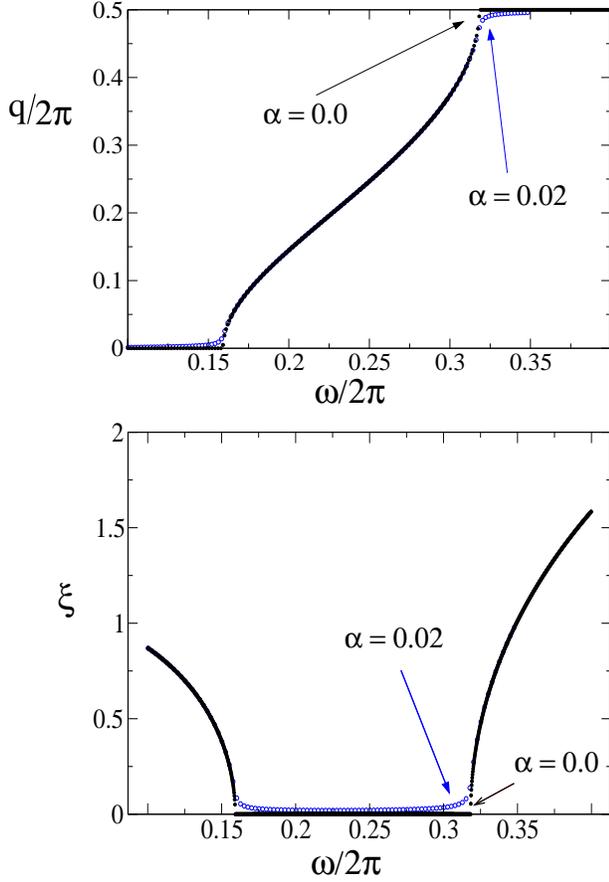}
    \caption{Wave vector $q$ and inverse of the decay length
      $\xi$ as functions of $\omega$ for two different values of the
      damping, $\alpha=0.02$ (open circles) and the Hamiltonian case
      $\alpha=0.0$ (filled ones).  The coupling parameter $C$ is in
      both cases equal to $0.75$.}
    \label{fig:xi_q-vs-w}
  \end{center}
\end{figure}

An example of the solutions $\xi(\omega)$ and $q(\omega)$ of
equations~(\ref{eq:dispersion}), for the particular values
$\alpha=0.02$ and $C=0.75$, appears in Fig.~\ref{fig:xi_q-vs-w}. For
comparison purposes, the graphs corresponding to the same value of
coupling for the Hamiltonian case are included.

\begin{figure}
  \begin{center}
    \includegraphics[width=0.45\textwidth]{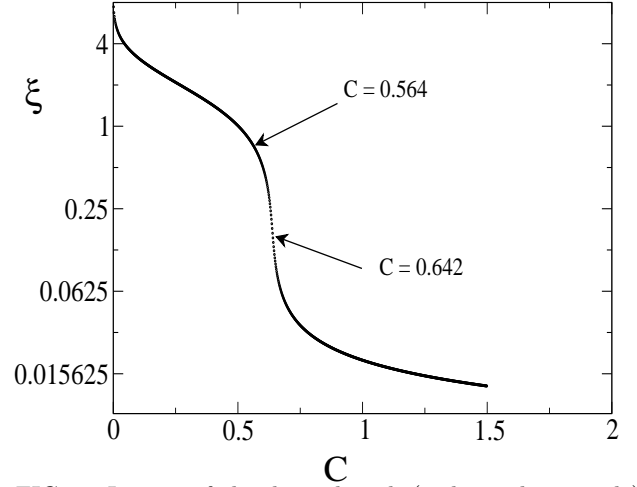}
    \caption{Inverse of the decay length (in logarithmic scale) as
      a function of the coupling parameter $C$, for a fixed value
      $\omega=3\omega_b$.  We mark two values of $C$, $0.564$ and
      $0.642$, which correspond to two rather different values of
      $\xi$, also used in the next figure.  The rest of the parameters
      are $\omega_b=0.2\pi$, $\alpha=0.02$ and $F_{ac}=0.02$.}
    \label{fig:xi-vs-c}
  \end{center}
\end{figure}

\begin{figure}
  \begin{center}
    \includegraphics[width=0.45\textwidth]{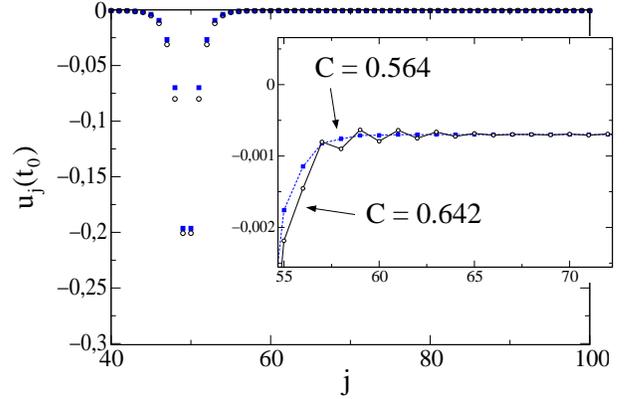}
    \caption{Breather profiles (two-site breathers) at a given time
      for two different values of $C$ ($0.564$ filled circles and
      $0.642$ open ones, see also the previous figure).  The inset
      shows the right hand side tails, where we can observe the
      existence of a phonon in the second case, corresponding to the
      entrance of the third harmonic of $\omega_b$ in the phonon band.
      The points are connected as a guide to the eye.}
    \label{fig:profilesfortwoC}
  \end{center}
\end{figure}

Note that for the existence of damped-forced discrete breathers there
is no need of a non-resonance condition (in contrast with the
Hamiltonian case), because for any frequency $\omega$, $\xi(\omega)
\neq 0$. However, for low values of $\alpha$, if some breather
harmonic $n\omega_b$ belong to the interval of values of $\omega$ for
which $\xi(\omega)$ is very small, the breather profile will show
large ``wings''.  In Fig.~\ref{fig:xi-vs-c} we plot $\xi(3\omega_b)$
as a function of the coupling $C$, for $\alpha = 0.02$ and $\omega_b =
0.2 \pi$.  Observe the dramatic decay of $\xi(3\omega_b)$ at around
$C=0.6$ corresponding to the entrance of the third breather harmonic
in the (so to speak) ``phonon band'', and compare the breather
profiles for two values of C, respectively below and above, in
Fig.~\ref{fig:profilesfortwoC}.  Both the wave-vector and the size of
the wings in figure fit very well with $q(3\omega_b)$ and
$\xi(3\omega_b)$ from equations~(\ref{eq:dispersion}).

\section{BIFURCATIONS AND PHASE DIAGRAM}
\label{sec:bifurcations}

We have continued numerically breather solutions from the uncoupled
(or anti-integrable) limit, for fixed values of the damping
coefficient ($\alpha=0.02$), external force frequency
($\omega_b=0.2\pi$) and intensity ($F_{ac}=0.02$).  The spectrum of
Floquet multipliers was also numerically computed for each solution,
thus monitoring their evolution on the complex plane.  The
configurations we have focused on are two: the one-site breather and
the two-site breather (adjacent sites).  Of course many other
configurations are possible, by choosing from all combinations of
sites in either the high-amplitude or low-amplitude attractor.
However these two simplest breathers already provide a quite rich
behavior, and, surprisingly, they allow continuation into very high
values of $C$, where the continuum limit is approached.

\begin{figure}
  \begin{center}
    \includegraphics[width=0.4\textwidth]{fig4a.eps}
    \includegraphics[width=0.45\textwidth]{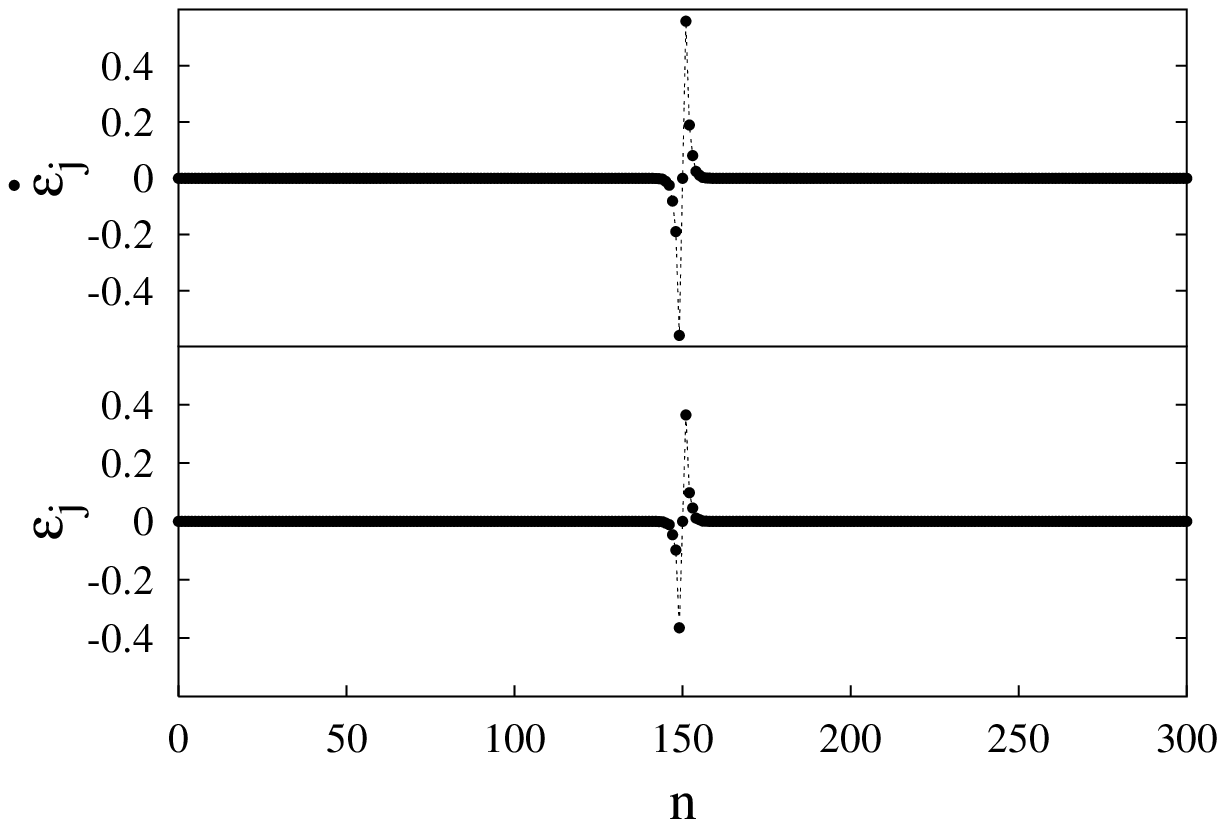}\\[0.5cm]
    \caption{Floquet spectrum for the one-site breather near
      $C=0.5296$ and closely after the first pitchfork bifurcation.
      All eigenvalues are in a circle of radius $\exp(-\alpha t_b/2)$
      except two, one of which crosses the unit circle by $+1$.  The
      lower figure shows the profile of the eigenvector corresponding
      to this unstable eigenvalue (both velocity and position
      components, $\dot\epsilon_i$, $\epsilon_i$).  Note that it is
      antisymmetric and strongly localized.}
    \label{fig:floquet-C1}
  \end{center}
\end{figure}

At very low values of $C$ the continued one-site breather is very
narrow, symmetric around its localization site ({\em i.e.} $u_{-i}(t)
= u_{i}(t)$ for all $t$ and $i$), and all the Floquet multipliers lie
on the circle of radius $\exp(-\alpha t_b/2)$ in the complex plane.
The breather remains stable for increasing coupling up to the value
$C_{\text{P1}}= 0.52962$ where a Floquet multiplier, which had
previously detached along the real axis from that inner circle,
reaches the unit circle at $+1$.  The corresponding eigenvector of the
Floquet matrix is localized around the breather site and possesses odd
mirror symmetry with respect to that site, as shown in
Fig.~\ref{fig:floquet-C1}.  Past the bifurcation, we are left with an
unstable symmetric breather and two new stable breathers, spatially
asymmetric and one being the mirror image of the other. We can
conclude that this is a {\em forward pitchfork
  bifurcation}~\cite{Str94}, associated to a spatial symmetry-breaking
transition of the discrete breather.

\begin{figure}
  \begin{center}
    \includegraphics[width=0.45\textwidth]{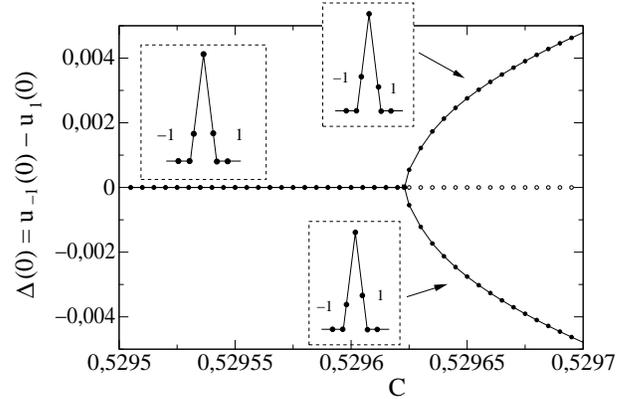}
    \caption{Profiles of the breather around the area of the 
      pitchfork bifurcation.}
    \label{fig:pitchfork}
  \end{center}
\end{figure}

In order to visualize the mirror symmetry breaking character of this
bifurcation, we plot in Fig.~\ref{fig:pitchfork} the difference
$\Delta(0)=u_{-1}(0)-u_{1}(0)$ at time $t=0 \pmod{t_b}$ of the
positions of the neighbor oscillators on both sides of the
localization site, as a function of the coupling parameter $C$ in the
vicinity of the bifurcation value. It is not surprising that, close to
the bifurcation, the difference $\Delta(0)$ scales with the coupling
parameter as $(C-C_{\text{P1}})^{1/2}$, because the one-dimensional
character of the unstable manifold of the symmetric breather allows to
reduce the analysis to that of a pitchfork bifurcation in a
one-dimensional map, where this scaling behavior for the distance
between branches is well-known.

The asymmetric stable branches born at $C_{\text{P1}}$ can be
continued for higher values of coupling.  We then observe how the
amplitude of one of the neighbors of the central site keeps
increasing, until it equals the amplitude of the central oscillator
(which has in turn decreased slightly), at $C_{\text{P2}}= 0.55315$.
It should be noted that the relative phases of the oscillators do not
appear to change, At this value, a {\em backward pitchfork
  bifurcation} occurs, where the two stable asymmetric breathers and
one unstable symmetric two-site breather merge, and a stable
mirror-symmetric {\em two-site} breather comes out.  As in the
bifurcation analyzed before, only one Floquet eigenvector, associated
to a Floquet multiplier of value $+1$, is involved.

This unstable two-site breather which joins this second pitchfork is
nothing but the two-site breather which can be continued from the
uncoupled limit.  Thus we have that the two elementary breathers
constructed at the uncoupled limit, one-site and two-site, undergo an
exchange of stability via this symmetry-breaking pitchfork mechanism.
For $C<C_{\text{P1}}$ the one-site breather is stable and the two-site
one unstable.  For $C_{\text{P1}}<C<C_{\text{P2}}$ both are unstable,
and the new asymmetric breather is stable.  Past $C_{\text{P2}}$, the
two-site breather is stable and the one-site unstable.  This is
exactly the same mechanism that was previously found by other
authors~\cite{AC98} for Hamiltonian breathers, and which is related to
breather mobility as we explore in the next section.  It is therefore
plausible to conjecture that the mechanism is highly generic and might
be expected in a large class of models.

To be thorough in our description, a much less interesting bifurcation
does appear in the two-site breather branch at very low $C$.  The
two-site breather is initially {\em stable} at the uncoupled limit,
but loses stability after a pitchfork bifurcation at $C\approx0.02$.
This is also a symmetry-breaking bifurcation as above, however it is
instructive to investigate the differences: this time the spatial
symmetry is broken in such a way that the new stable, asymmetric
breathers suffer a {\em de-phasing} between the two central sites.
This can be confirmed rigorously by examination of the relevant
eigenvector at the bifurcation.  If one takes as reference for the
time origin the instant at which the two central sites have maximum
amplitude, the unstable eigenvector for this second pitchfork shows
components only in the velocity part, not in amplitudes; the former
bifurcation shows exactly the opposite behavior.  In any case, the
continuation of the asymmetric breather from this bifurcation at
$C=0.02$ is quickly lost after a Hopf bifurcation, and we have not
found any more interesting behavior arising from these curious
branches.

\begin{figure}
  \begin{center}
    \includegraphics[width=0.4\textwidth]{fig6a.eps}
    \includegraphics[width=0.45\textwidth]{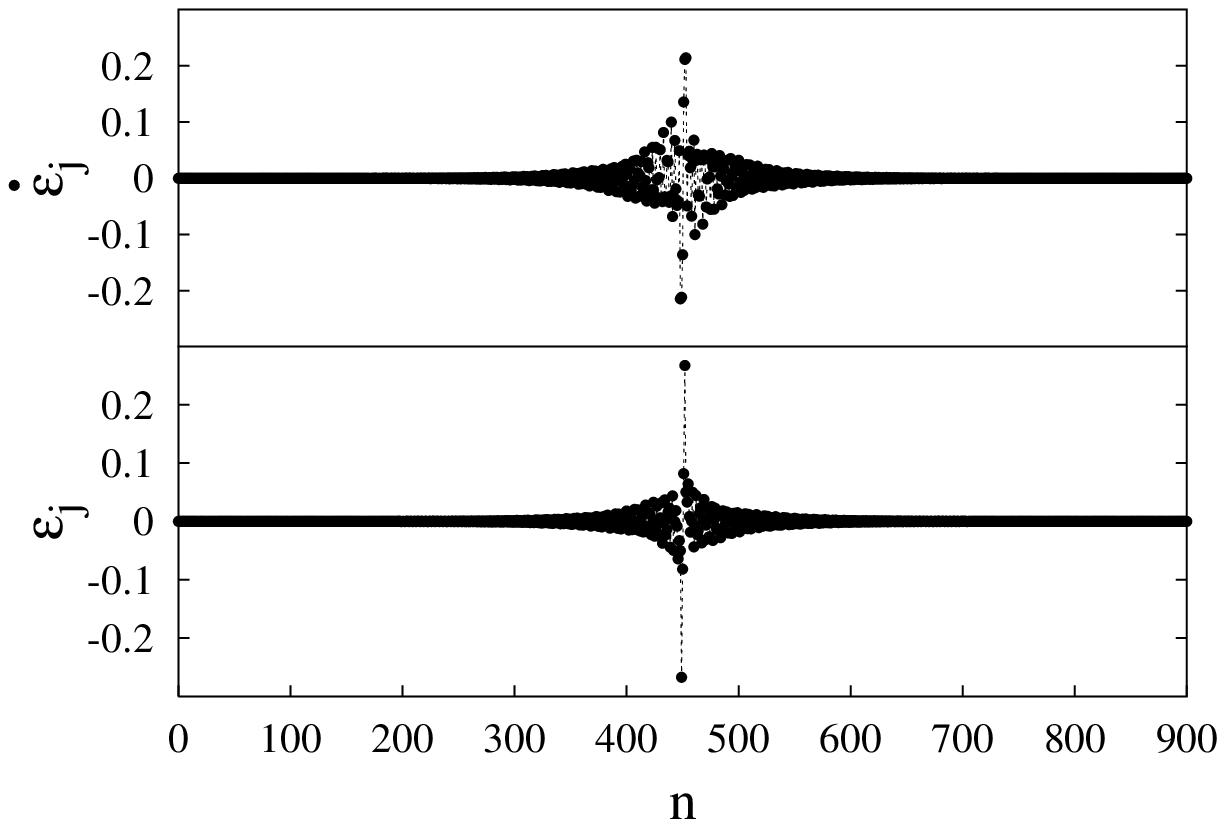}\\[0.5cm]
    \caption{Floquet spectrum for the Hopf bifurcation at
      $C=0.871$.  An eigenvalue and its complex conjugate cross the
      unit circle at a non-zero angle in the complex plane.  The
      figure at the bottom shows the corresponding eigenvector.}
    \label{fig:floquet-Hopf}
  \end{center}
\end{figure}

Now we turn on to the continuation of the stable two-site breather
branch past $C>C_{\text{P2}}$.  We find now that near $C=0.817$ two
complex conjugate Floquet multipliers cross the unit circle at
$\exp(\pm i\varphi)$, with $\varphi=1.285$, and the (periodic)
two-site breather becomes unstable.  The real and imaginary components
of the associated eigenvectors are shown in
Fig.~\ref{fig:floquet-Hopf}.  Close to the bifurcation, small
perturbations of the unstable periodic breather bring it into a
quasiperiodic breather (as verified by inspection of the Poincar\'e
section), thus confirming the scenario of an {\em (Andronov-) Hopf}
bifurcation~\cite{KH95}.  Indeed, the power spectrum analysis of the
quasiperiodic attractor (see figure) reveals two basic frequencies,
$\omega_b$ and $\omega_{\text{new}}=0.0873$.  However, since this new
frequency is rather different from the frequency
$\varphi\omega_b/2\pi= 0.1285$ associated to the destabilizing
eigenvalue couple, one concludes that the stable quasiperiodic
attractor does not come out straight from the bifurcation.  In other
words, the simplest scenario of a {\em subcritical} Hopf bifurcation
is discarded.  Moreover, the quasiperiodic attractor can be easily
continued back for lower values of the coupling ({\em i.e.\/},
$C<0.817$), therefore confirming that we have a {\em subcritical} Hopf
bifurcation.  Note that, as parameters other than $C$ change, it is
possible for these subcritical Hopf bifurcations to become
supercritical, for their genericity lies in the Hopf character, not in
being subcritical or supercritical.

The lack of periodicity for the new (quasiperiodic) breather attractor
prevents the use of Floquet analysis.  However, its stable character
can be numerically ascertained, checking its robustness against small
perturbations in the dynamics.  This quasiperiodic two-site breather
turns out to be stable for couplings up to $C=0.88$, beyond which it
starts moving spontaneously.  Only when we reach $C=0.96$ we recover
again a stable, pinned quasiperiodic breather.  Meanwhile, the pinned,
periodic two-site breather (which became unstable after the Hopf
bifurcation at $C=0.817$) can be continued by a Newton method.  At a
value near $C\approx0.995$, it rejoins the quasiperiodic two-site
breather in an inverse (now supercritical) Hopf bifurcation, becoming
stable again.

\begin{figure}
  \begin{center}
    \includegraphics[width=0.45\textwidth]{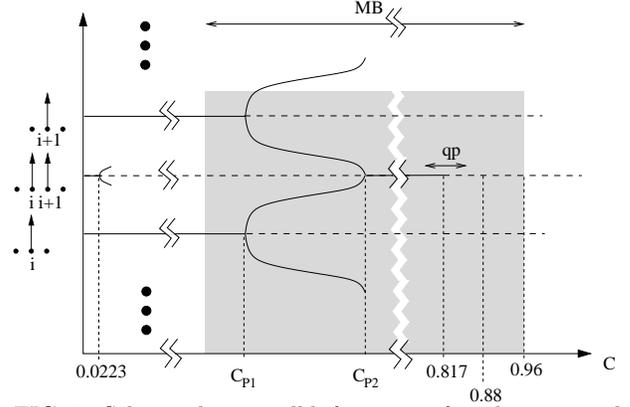}
    \caption{Scheme showing all bifurcations found in our model up to
      values of $C\approx 1$.  The most relevant ones are probably the
      pitchforks which exchange stability between the one-site and the
      two-site breather ($C_{\text{P1}}$, $C_{\text{P2}}$).  At
      $C=0.817$ the two-site breather has a subcritical Hopf
      bifurcation, connecting it to a quasiperiodic breather.  At
      $C=0.88$ this quasiperiodic solution disappears as it turns into
      a slow moving breather.  The grey-shaded areas between $C=0.51$
      and $C=0.96$ are those where moving breathers, either fast or
      slow, can be found.}
    \label{fig:allbifurcations}
  \end{center}
\end{figure}

We have tried to summarize most of these bifurcations in the sketch of
Fig.~\ref{fig:allbifurcations}.  We postpone to the next section the
analysis of the observed mobile breathers in this region (and others)
of parameter space.

To conclude this section we will comment on the continuation and
stability analysis as $C\rightarrow\infty$, {\em i.e.\/} the so-called
continuum limit.  The first interesting fact is that both the one-site
and two-site breather, whether stable or unstable, have been found to
be continuable for couplings as high as desired.  In other words, they
never disappear by, say, saddle-node bifurcations or the like.
Another interesting point is that for $C>1$ we have also found
pitchforks which connect the two branches, exactly in the same way as
the first one at $C_{\text{P1}}$, $C_{\text{P2}}$.  The ranges of
coupling between forward and backward pitchfork bifurcations (that is,
where the connecting branches of stable asymmetric breathers exist)
get progressively narrower for higher coupling values.  And finally,
both the breather profiles and their Floquet spectrum reveal a very
natural approach to the continuum limit: the solutions get broader in
size, while the eigenvalue responsible for the pitchforks remains
closer and closer to $+1$ at all times, announcing the appearance of
the Goldstone mode (due to translational invariance) as
$C\rightarrow\infty$.  These effects were ostensibly manifest at
$C\gtrsim5$.

\section{MOBILE BREATHERS}
\label{sec:mobile}

The problem of the mobility of discrete breathers is still very poorly
understood. While the fundamental theory of stationary breathers is
now firmly established~\cite{MA94}, moving discrete breathers (MB for
short) have so far eluded a rigorous treatment.  But the fact is that
moving breathers have been observed and studied through numerical
simulation in various works~\cite{SPS92,HT92a,HT92b,AC98,FK99}, and
they appear to be a phenomenon with the same degree of genericity as
the stationary case.  Up to now most of those works have dealt with
Hamiltonian systems; here we present a study of moving breathers in
our forced and damped F-K model.  Some of the results are strikingly
similar to those observed in Hamiltonian systems.

We should first point out that a moving breather is something to be
distinguished from a similar class of solutions, namely {\em lattice
  solitons}~\cite{EF90,DEFW93}.  In a lattice soliton a pulse
propagates without dispersion through the lattice, but, unlike
breathers, there is no ``internal'' oscillation.  This additional
degree of freedom makes the moving breather a more complicated object.
For instance, in Hamiltonian lattices, it is easy to see that
inevitably the moving breather resonates with the phonon band (note
the presence of a quasi-periodic spectrum due to the additional
frequency introduced by the translational motion), and therefore it is
not possible to have tails which decay to zero.  It is not clear
whether the solution is just a transient which eventually decays by
phonon radiation, or maybe an infinite lifetime breather which
``rides'' on an infinite, small amplitude radiation background.

But our model here is dissipative and has external forcing, and it
turns out that moving breathers appear as proper attractors of the
dynamics.  Since these solutions are not transients, we can study and
characterize them accurately and with great confidence.  Even though
this will not shed any light on the problem of existence of
Hamiltonian moving breathers, there is another aspect of theoretical
interest in which this study can contribute: the (possible) concept of
a Peierls-Nabarro barrier for breather motion.  This concept arises
because of the similarities with the problem of mobility of
discommensurations (kinks) in the Frenkel-Kontorova model~\cite{FM96}.
The discommensuration is an equilibrium static structure, for which
one may ask how much energy it costs to displace it by one lattice
site, until it reaches the equivalent configuration by (discrete)
translational invariance.  This is commonly referred to as the
Peierls-Nabarro (PN) barrier.  And it is possible to give a precise
definition: from all possible continuous deformations of the initial
configuration into the final one, take the one in which the maximal
energy change along the path is the minimum.  Very fruitful results in
the theory of the F-K model have stemmed from this
definition~\cite{FM96}.

But a corresponding definition of a PN barrier for breathers proves
quite problematic.  The difficulty lies in that it is not clear which
space to use, since the configurations are now periodic functions, not
static points.  Some authors have suggested possible candidates for a
rigorous definition, but the issue is still under
debate~\cite{FW94,Aub97,AMS99}.  Technicalities aside, it is still
possible to give a working definition of the Peierls-Nabarro barrier
for breathers, at least in some cases.  Most studies generate moving
breathers by perturbing stationary ones, and Ref.~\cite{CAT96} gave a
systematic method to do this.  Looking at the linear stability
analysis of the breather (Floquet analysis), they found that in many
cases one can identify an eigenmode which is distinctively localized
and whose spatial symmetry is the appropriate for breather motion
(note that similar depinning modes are responsible for the depinning
of discommensurations~\cite{FM96} under uniform forcing).  It was
found that adding a perturbation along this depinning eigenmode,
provided one overcomes a certain threshold, results in a moving
breather.  Such thresholds are probably the best pragmatic approach to
the definition of a PN barrier.

A further study~\cite{AC98} showed that many Hamiltonian lattices
exhibit a very interesting behavior which is linked to mobility.  It
was found that, as the coupling is increased from $C=0$, the one-site
breather and the two-site breather each undergo a pitchfork
bifurcation, where new branches of periodic but spatially asymmetric
breathers emerge.  These branches do in fact connect those pitchfork
points, and the corresponding Floquet eigenmodes responsible for the
bifurcations obviously show a spatial symmetry which we could dub as
``depinning''.  This is exactly what we have found in our dissipative
model, as shown in the previous section.  And, just as in the cited
works on Hamiltonian systems, we have also verified that the mobility
is greatly enhanced for values of coupling in the vicinity of these
bifurcations: very small amounts of perturbation along the depinning
mode are enough to create the mobile breather.  The upshot is that
this phenomenon provides a mechanism for the existence of mobile
breathers at relatively low couplings (high discreteness) and with
very slow velocities, two properties which were counterintuitive and
unexpected.

We should note that the continuous sine-Gordon equation under external
ac forcing and losses does not support MB solutions \cite{QS98}. The
way in which a continuous breather destabilizes is by a transition to
a quasiperiodic state and finally creation of a kink-antikink pair
\cite{LS88}.

In the following we begin exploring this relation between the
stability of stationary breathers and the existence of their mobile
counterparts. Then we concentrate on studying the properties of moving
breathers in dissipative systems in more detail.  Finally, we explore
other aspects such as collisions.

\subsection{Generation and phase diagram of MB}

\begin{figure}
  \begin{center}
    \includegraphics[angle=270,width=0.45\textwidth]{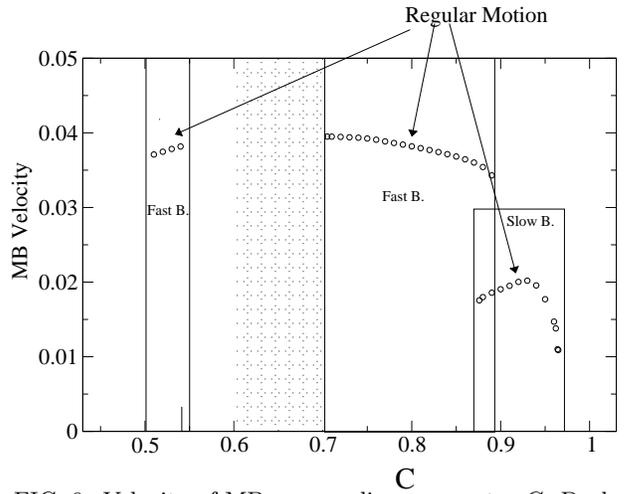}
    \caption{Velocity of MB vs.\ coupling parameter $C$. Dashed area shows
      the region in which MB have a diffusive motion. It is also
      showed different regions of ``slow'' and ``fast'' MB. See the
      text for details. }
    \label{fig:diagr_faseBM}
  \end{center}
\end{figure}

We have found MB as proper attractors of the dynamics in a wide range
of couplings, in particular for $0.5\lesssim C\lesssim0.96$.  We have
found them either by excitation of the depinning mode of stationary
breathers (as explained below) or simply by letting the system evolve
to a steady state after the instabilities of some stationary breathers
develop fully.  Then it is possible to carry out a continuation of the
MB into other parameter values, since their attractor property allows
to change slowly a parameter and track the MB solution.
Figure~\ref{fig:diagr_faseBM} shows the phase diagram of MB we have
constructed with this procedure.

\begin{figure}
  \begin{center}
    \includegraphics[width=0.45\textwidth]{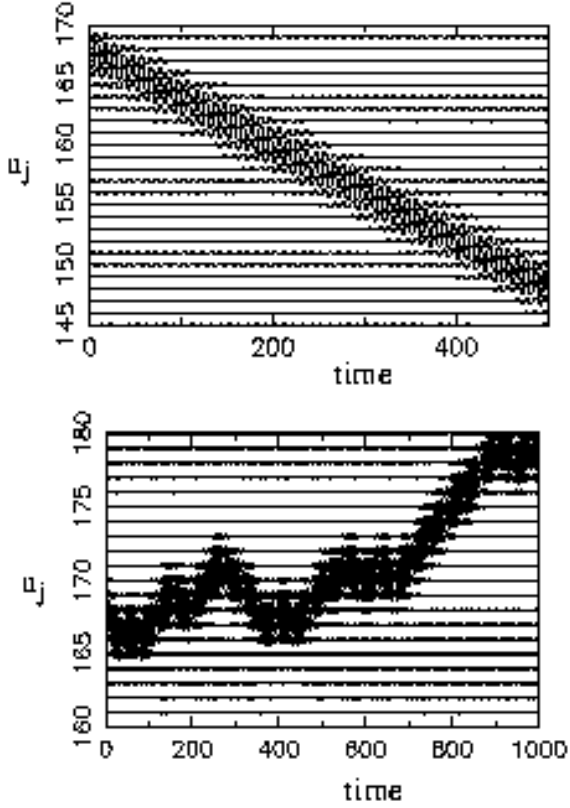}
    \caption{The upper figure shows a MB with regular motion at $C= 0.75$.
      Below, a MB with diffusive motion at $C=0.65$.}
    \label{fig:mobile_breathers}
  \end{center}
\end{figure}

Around the first symmetry-breaking bifurcation there is a narrow
region in which we found MBs with regular motion and well defined
velocity (Fig.~\ref{fig:mobile_breathers}).  A similar region is found
in the interval $0.7\lesssim C\lesssim 0.88$. We call these solutions
{\em induced fast breathers}.  To generate these {\em steady state} MB
we have followed the procedure described in~\cite{CAT96}.
Surprisingly, this method works very well not only near the
bifurcations.  We typically use as initial conditions:
\begin{equation}
  u_i(t=0) = u^{0}_i + \lambda \epsilon^a_i ,
  \label{eq:perturb}
\end{equation}
where $u_i^{0}$ is a stationary breather solution, $\epsilon^a_i$
corresponds to the antisymmetric (and localized) eigenvector mode at
the bifurcation taking place near $C=0.53$, and finally $\lambda$
measures the strength of the perturbation applied.  It is found that,
as in the Hamiltonian case, a critical $\lambda_c$ is necessary to
unpin the breather.  However, in contrast to the results for
Hamiltonian systems, once the breather starts to move the velocity is
unique (independent of $\lambda$), i.e. the MB is a robust attractor
of the dynamics. Once a {\em induced} MB is generated at a value of
$C$, this solution can be continued by varying $C$ slowly, with almost
no variation in velocity.  The analysis of Poincar\'e sections of this
MB shows clearly a quasiperiodic behavior.  Therefore there is {\em
  not commensurability} between the internal frequency $\omega_b$ and
the new frequency associated to the velocity $\omega_{\text{mb}} = 2
\pi v_{\text{mb}}$. This behavior precludes the use of fixed point
methods (like Newton method) based in the periodicity of solutions, to
find (numerically) exact MB solutions. Also, this quasiperiodicity
prevent us to extend the Floquet analysis to MB.

For intermediate couplings $0.60\lesssim C\lesssim0.72$ (shadow region
in the phase diagram), the breather portrays random motion.  For some
time interval, the MB moves regularly in one direction, then suddenly
remains inmobile (but quasiperiodic) for a while, changing then its
motion to the other direction, and so on.  A plausible scenario is
that of the occurrence of a \emph{crisis}~\cite{Ott93} at
$C\approx0.72$ which destabilizes the regularly moving breathers (of
positive and negative velocity, respectively), giving rise to a
chaotic attractor consisting of intervals (of random length) of
approximately regular motion, followed by changes of direction.

As we have already mentioned, at $C=0.88$ the stationary breather
solution (quasiperiodic) disappears, and only the MB solutions
survive.  We will refer to this breather as {\em spontaneous slow} MB.
Its velocity is approximately half of the {\em fast MB} and shows a
great variation as a function of $C$.  Moreover, there is a narrow
window around $C=0.89$ where both the slow and fast MB coexist.

\begin{figure}
  \begin{center}
    \includegraphics[width=0.45\textwidth]{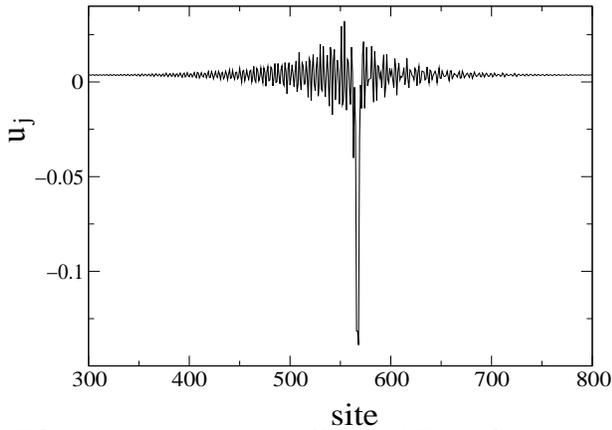}
    \caption{Instantaneous profile of a MB at $C = 0.75$ and
      $F_{ac}= 0.045$.  Note the clear asymmetry of the phonon tails
      in front of and behind the breather.}
    \label{fig:basym}
  \end{center}
\end{figure}

Regular motion MB are also stable against variation of parameters
other than $C$.  By increasing $F_{ac}$, the velocity of the breather
increases, showing a very asymmetric profile, as shown in
Fig.~\ref{fig:basym}.  As we explain below, the origin of this
asymmetric shape can be explained in terms of forward and backward
emission of phonons from the moving breather, which suffer a Doppler
effect.  Above a critical value of $F_{ac}$, which is dependent on
$C$, a shock wave is formed and the regular motion becomes again
diffusive.

\subsection{Emission of phonons}

An important feature of both {\em pinned quasiperiodic} and {\em
  moving} breathers is the emission of low amplitude linear waves
(phonons).  In both cases the breather tails clearly show a complex
quasiperiodic behavior in which many frequencies are involved.  For
the moving breather, these tails are also markedly asymmetric, due to
the translational motion.  Other authors have investigated the
behavior of Hamiltonian breathers when subject to phonon
scattering~\cite{CAF98}; note that in our case it is the breather
itself the source of phonons.

In order to investigate the phonon emission we have computed the power
spectrum of $\dot{u}_i(t)$ for sites sufficiently far away from the
breather center, as given by the expression
\begin{equation}
    S(\omega) =  \left| \int_{-\infty}^{\infty} \dot{u}_j(t)
      e^{i\omega t} dt \right| ^2
  \label{eq:PowerS}
\end{equation}
In all spectra, we can observe peaks corresponding to the driving
frequency and its odd harmonics, as expected.  We also observe a broad
band spectrum corresponding to frequencies in the phonon band, with
several resonant peaks.  In both cases we can explain those
frequencies satisfactorily in terms of emission of phonons by the
breather.

\begin{figure}
  \begin{center}
    \includegraphics[width=0.45\textwidth]{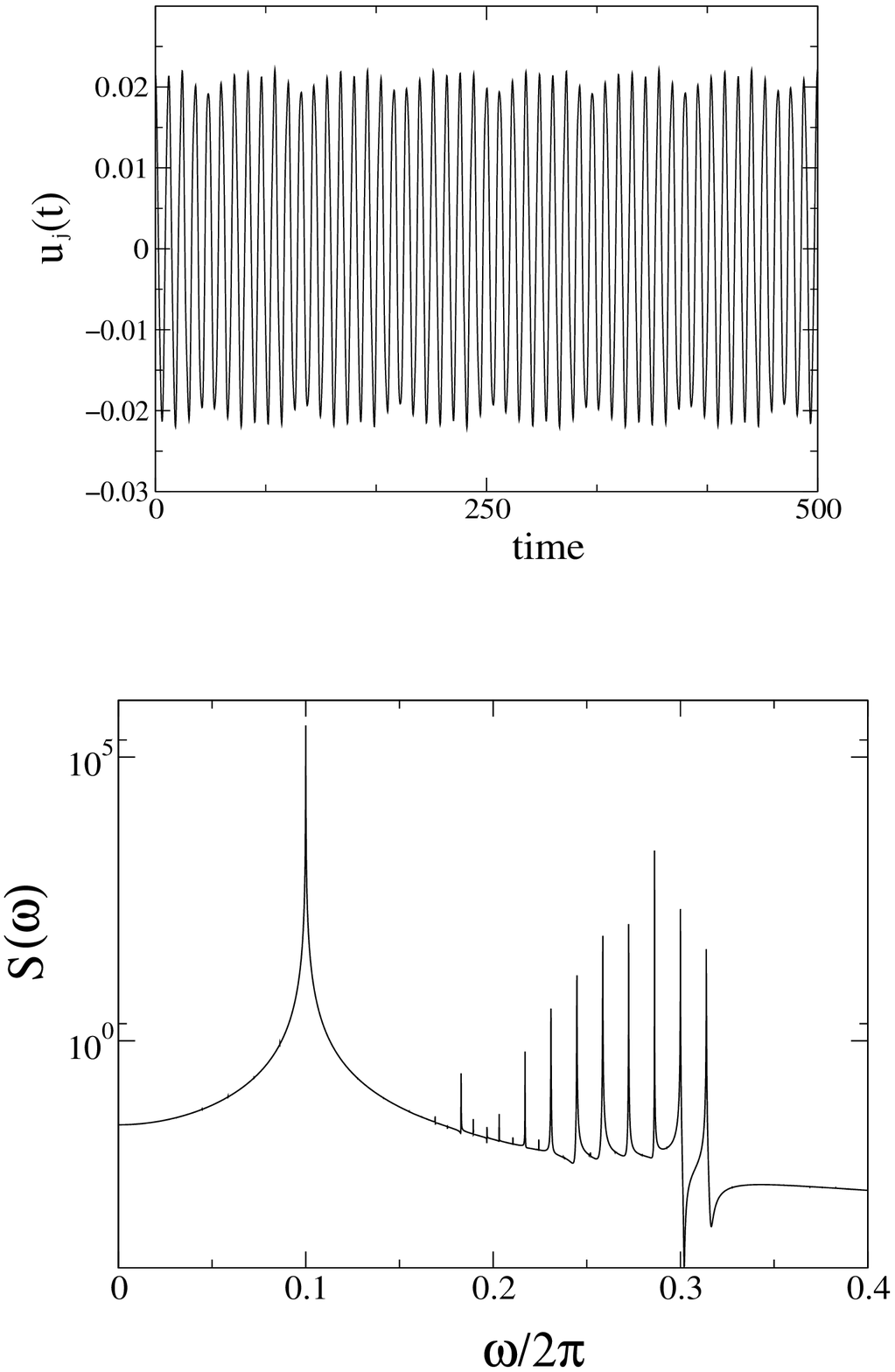}
    \caption{Time evolution for the tail of quasiperiodic breather 
      at $C=0.817$ (upper figure) and the corresponding power spectrum
      (down figure) whose peaks are given by linear combinations of
      two frequencies $\omega_b$ and $\omega_{new}$.}
    \label{fig:phonon_qp}
  \end{center}
\end{figure}

Figure~\ref{fig:phonon_qp} shows a time snapshot and the corresponding
power spectra for a particle in the tails of a quasiperiodic pinned
breather.  In this case, the resonant peaks simply correspond to
frequencies which are linear combinations of $\omega_b$ and
$\omega_{\text{new}}$ (the second basic frequency of the quasiperiodic
breather):
\begin{equation}
  \omega_{\text{tail}} = m \omega_b + n \omega_{\text{new}}, 
  \qquad m,n \in {\Bbb Z} 
  \label{eq:linearcomb}
\end{equation}

\begin{figure}
  \begin{center}
    \includegraphics[width=0.45\textwidth]{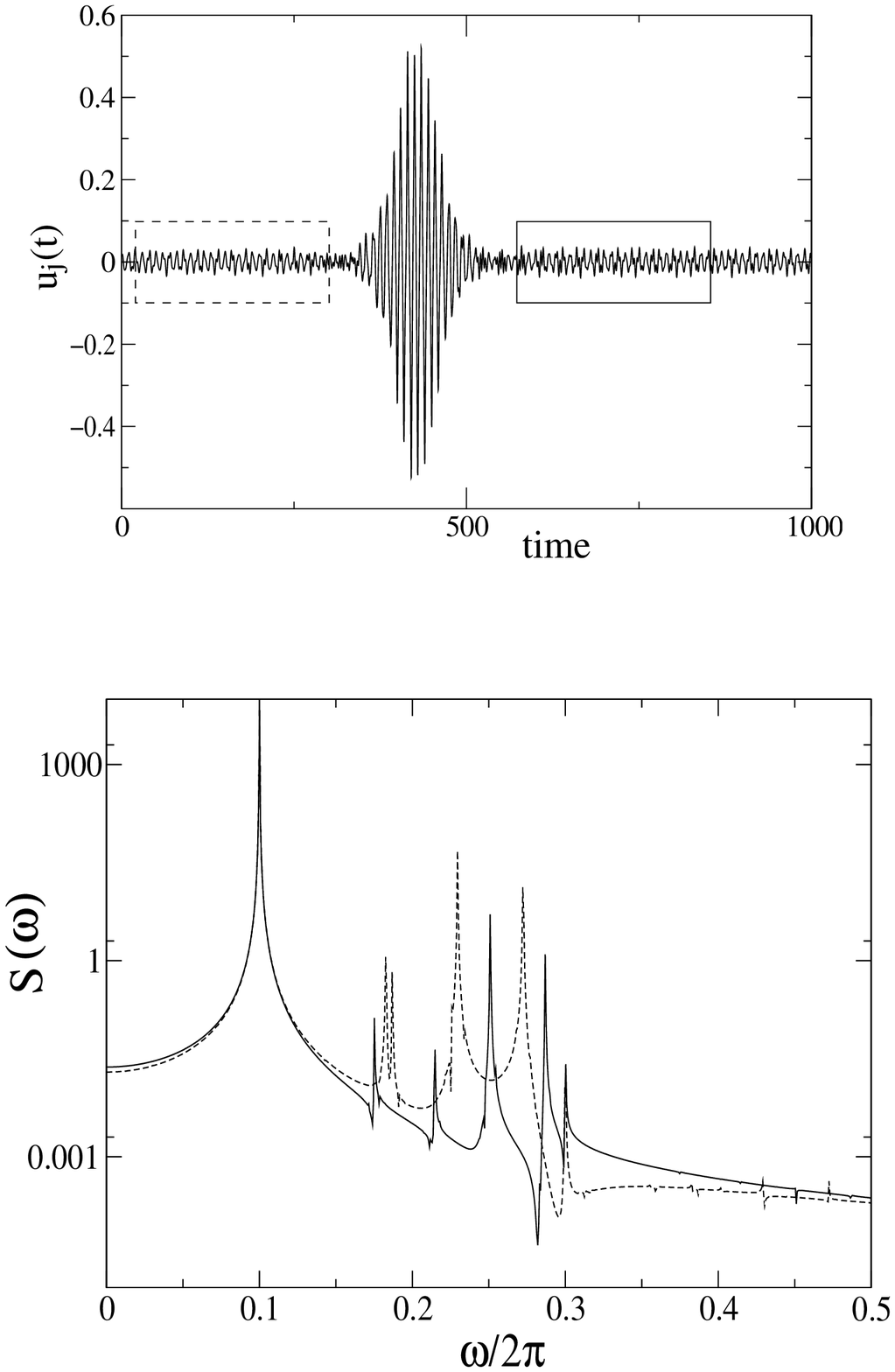}
    \caption{Time evolution for a particle which is passed by a MB
      (upper).  Below, power spectra corresponding to the tails before
      and after the passage of the breather.  Note how the peaks are
      shifted by Doppler effect.}
    \label{fig:doppler}
  \end{center}
\end{figure}

In the case of moving breathers, the peaks also correspond to the
frequencies given above, but shifted by the Doppler effect since they
are emitted by a moving source.  Their calculation requires thus a
little more care.  We recall that the propagation of phonons is given
by the dispersion relations (\ref{eq:dispersion}).  The frequency of
the emitting source, in the reference frame of the source itself, is
$\omega_{\text{src}}=n\omega_b+2\pi m v_{\text{mb}}$.  This second
frequency appears because the breather is moving over a periodic
potential with velocity $v_{\text{mb}}$.  However, in the reference
frame of the medium (the lattice), this frequency will be modified
according to:
\begin{equation}
  \omega_{\text{tail}} = \omega_{\text{src}}
    \pm 2 \pi q(\omega_{\text{tail}}) v_{\text{mb}},
  \label{eq:Doppler}
\end{equation}
which is the well-known Doppler effect, only that the medium is
dispersive~\cite{Lan82}.  In particular, note how the wavevector $q$
of the propagated phonon depends on $\omega$, as given by
Eqs.~(\ref{eq:dispersion}).  Therefore Eqs.~(\ref{eq:dispersion}) and
(\ref{eq:Doppler}) have to be solved self-consistently for $q$ and
$\omega_{\text{tail}}$, and then the different peaks of the power
spectrum can be worked out.  The agreement of this calculation with
the observed frequencies (see Fig.~\ref{fig:doppler}) is excellent.

Finally we remark that for small lattice sizes, and when using
periodic boundary conditions, tails in front of and behind the
breather can overlap.  In such cases, solutions are similar to the
so-called {\em nanopterons}~\cite{Boy90} of Hamiltonian systems, in
which the MB appears to move in a ``sea'' of phonons.

\subsection{Breather Collisions and Thermal Effects}

\begin{figure}
  \begin{center}
    \includegraphics[width=0.45\textwidth]{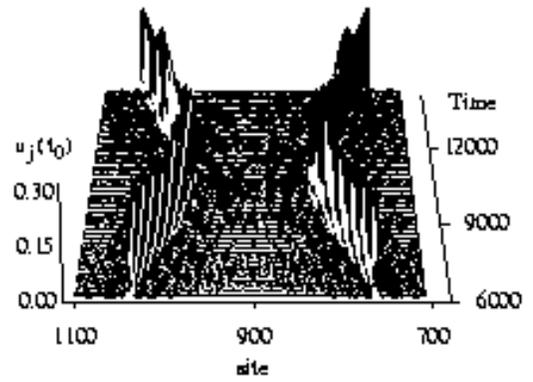}
    \caption{An elastic collision between two slowly moving
      breathers.  It is possible to appreciate how the breathers
      ``see'' each other through their phonon tails.  This collision
      was obtained at $C=0.89$.}
    \label{fig:3dcollision}
  \end{center}
\end{figure}

Since we are dealing with a system in which pinned (periodic and
quasiperiodic) and mobile breathers coexist for the same parameter
values, it seems natural to study their stability against collisions
between them.  Very different kind of events appear depending on the
breather velocity and the initial conditions (initial distance between
breathers).  The faster the breathers are, the more likely they are to
destroy each other.  The main result is that, when the breathers
survive to the collision, the interaction is mediated by the phonons
which dress the breather.  For large velocity, the tail in front of
the breather is short (the MB is very asymmetric), so the breather
cores can overlap and we observe they get destroyed.  For moderate
velocities, the opposite occurs: the tails are large, and it seems as
if they mediated the collision, slowing down the breathers and
preventing the cores to touch.  A clear example of this latter case
can be seen in the figure~\ref{fig:3dcollision}, where we observe an
``elastic'' collision in which the breathers approach each other up to
a distance which is the range of the phonon tail. Note that this
distance ($\approx 150$) is much larger than the core breather size
($\approx 10$).

\begin{figure}
  \begin{center}
    \includegraphics[height=0.80\textheight]{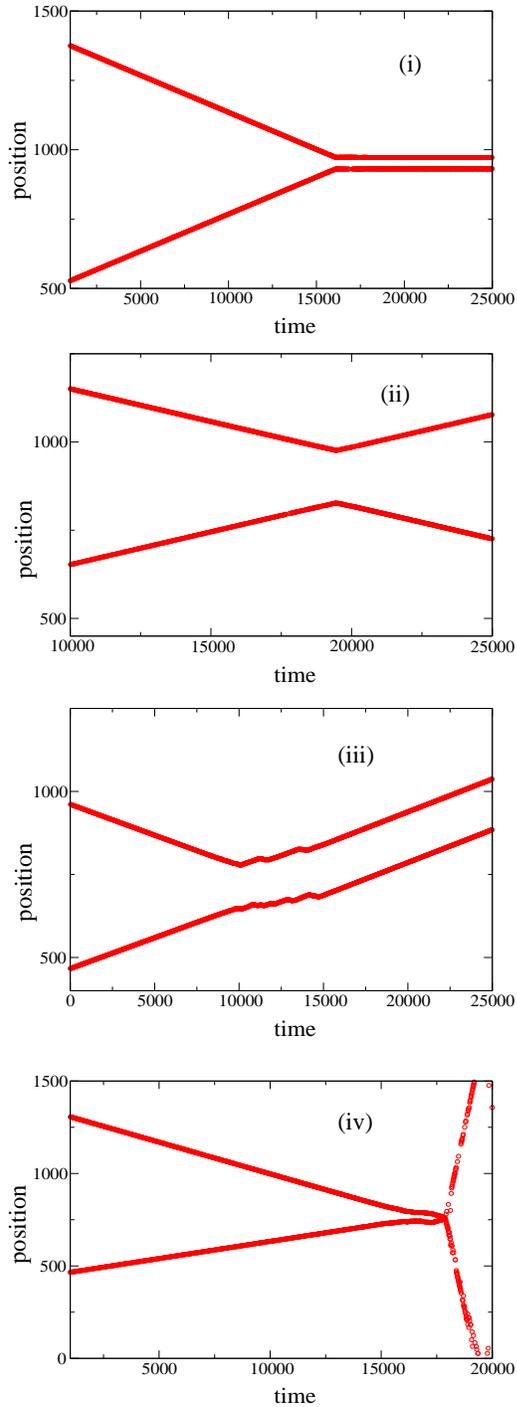}
    \caption{Four scenarios of collisions between breathers. We represent
      the traces by plotting the position of the energy maximum. 
      (i) Formation of a
      pinned ``molecule'' after a collision of two MB at $C=0.75$ and
      $F_{ac}=0.011$.  (ii) ``elastic'' collision at $C=0.89$ and
      $F_{ac}=0.02$. (iii) Formation of mobile ``molecule'' $C=0.89$
      and $F_{ac}=0.02$. (iv) Annihilation of a ``slow'' and a
      ``fast'' breather at $C=0.89$ and $F_{ac}=0.02$. In this last
      case, the traces after the collision correspond to linear
      radiation (phonons).} \label{fig:collisions}
  \end{center}
\end{figure}

In figure~\ref{fig:collisions} we show some of the multiple behaviors
we have observed in simulations: {\em (i)} Collision between two MB at
low $F_{ac}$ and thus low velocity, at $C=0.75$. Their velocity is low
enough to not destroy themselves, but to create a new multibreather
state; {\em (ii)} elastic collision of two ``slow'' breathers at
$C=0.89$; {\em (iii)} collision which gives a {\em mobile}
two-breather state.  The last two collisions only differ in the
initial conditions.  Finally, {\em (iv)} breather annihilation between
a ``slow'' and ``fast'' breather at $C=0.89$ where they coexist.

An interesting phenomenon arising from the simulations above is the
formation of multibreather solution both pinned and mobile. These
breather ``molecules'' are appear to be linked by ``phonon bonds''.
Surprisingly, these multibreather solutions are more robust against
changes of parameters.  For instance, they can be subject to larger
$F_{ac}$ without losing regular motion and then reaching larger
velocities.  However, a systematic study and rigorous characterization
of these configurations will be left for further publication.

Finally, we have incorporated in the equations of motion
(\ref{eq:mot}) a random force $\xi_i(t)$ with $<\xi_i(t)> = 0 $ and
$<\xi_i(t) \xi_j(t')> = 2T \delta_{i,j} \delta(t-t')$ in order to
simulate the Langevin dynamics of our system. For low $T$ ($T <
10^{-4}$) breathers solutions (pinned and mobile) are stable in the
whole range of coupling $C$, in the sense that localization persists.
To be precise, we observe that, if we are in a region of parameters
where the MB exists, the noise always induces motion, which is of
course stochastic itself. This can be understood again in a scenario
of crisis induced by the thermal noise.  For higher $T$ the thermal
excitation of kink-antikink pairs and other breathers masks the
original breather.

\section{DISCUSSION and CONCLUSIONS}
\label{sec:conclusions}

Dissipative discrete breathers (DB's, for short) are generic solutions
of forced-damped lattices of non linear oscillators. Contrary to their
Hamiltonian counterparts, which are severely affected by harmonic
resonances with the phonon band, the intrinsic localization of energy
in the dissipative case is not easily destroyed by resonances due to
the efficient damping of the radiation away from the localization
site. The character of attractor of the dissipative DB's allows their
numerical continuation in parameter space with simpler procedures than
those needed for the continuation of Hamiltonian DB's, and their
robustness against all kind of small perturbations (including
stochastic ones) ensures their observability in experimental
situations.

Pinned dissipative DB's which are continued from the uncoupled limit
experience generically different kinds of instabilities by localized
modes, namely pitchfork (forward and backward) and Hopf (supercritical
and subcritical) bifurcations.  Pitchfork bifurcations produce DB's
with broken mirror symmetry, while Hopf bifurcations lead to
quasiperiodic DB's.  In any case it is remarkable that the two basic
solutions (one-site and two-site periodic breather) are found to be
continuable (as stable or unstable solutions) up to the continuum
limit, where continuous translational invariance is restored in the
model and then a Goldstone mode appears in their Floquet spectrum.
For not-too-large forcing, tail analysis as explained in
section~\ref{subsec:tail} explains successfully the numerical power
spectra in sites away from the breather center, as well as DB
profiles.

In certain regions of the parameter space, mobile DB's occur as
attractors for an open set of initial conditions (basin of attraction)
in phase space.  Their velocity is determined by the model parameters,
and it is slow compared with the time scale set by the forcing
frequency.  One class of mobile solutions are connected to the
existence of depinning modes in the Floquet spectrum of periodic DB's,
when they are in the vicinity of (symmetry-breaking) pitchfork
bifurcations.  This mechanism allows the generation of mobile DB's for
values of the coupling which are surprisingly low.  Another class is
related to the De-stabilization of quasiperiodic pinned breathers, and
their velocity is even slower than that of the previous class.

Finally, we tested the robustness of breathers by means of collisions
and the application of stochastic perturbations (Langevin noise).  It
is concluded that breathers are quite robust in the sense that the
localization persists; however, in regions of parameters where mobile
DB's exist, the thermal noise induces random motion in the breather.

\section{ACKNOWLEDGEMENTS}
We acknowledge C.~Baesens, A.~S\'anchez, and J.~J.~Mazo for many
useful discussions on this work.  Financial support is acknowledged to
DGES PB98-1592 of Spain, Acci\'on Integrada Hispano-Brit\'anica
HB1999-0104, and European Network LOCNET HPRN-CT-1999-00163.  JLM
acknowledges a Return Grant from the Spanish MEC.

%\bibliographystyle{prsty}
%\bibliography{breatherfk}
%\bibliography{publish,journal,mybibfile}

\ifx\undefined\allcaps\def\allcaps#1{#1}\fi\ifx\undefined\allcaps\def\allcaps#%
1{#1}\fi

%%% Template for fig.  The actual inclusion of the fig has to be
%%% removed before sending the compuscript via e-mail.
%\begin{figure}
%  \begin{center}
%    \includegraphics[width=0.5\textwidth]{figs/???.eps}
%    \caption{Bla bla}
%    \label{fig:dummy}
%  \end{center}
%\end{figure}

\begin{figure}
  \begin{center}
    \includegraphics[width=0.45\textwidth]{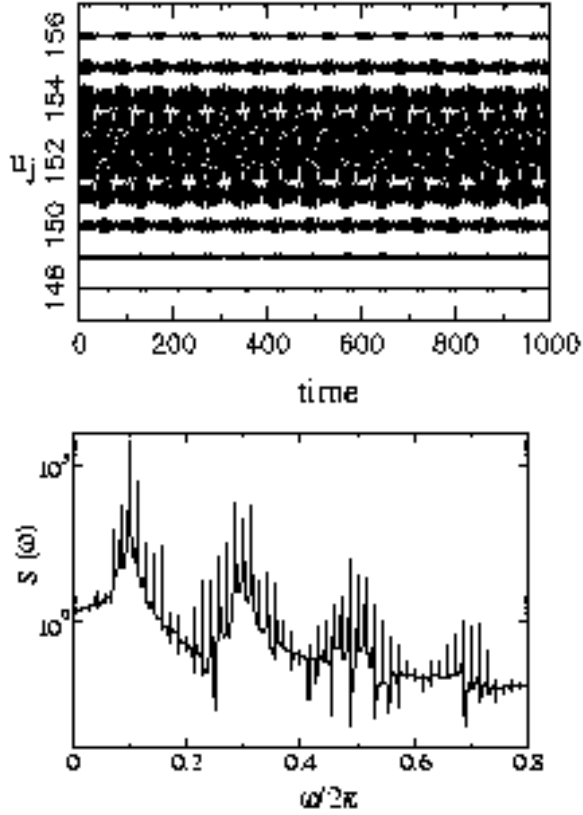}\\[1cm]
%    \includegraphics[angle=270,width=0.45\textwidth]
%    {fig7b}\\[0.5cm]
    \caption{Two-site quasiperiodic breather.  Note that
      particles on both sides of the breather are out of phase. The
      figure below shows the power spectrum of one of the central
      particles.  The peaks are linear combinations of the two
      relevant frequencies, $\omega_b$ and $\omega_{\text{new}}$.
      Note that only odd harmonics of $\omega_b$ appear.}
    \label{fig:2s-qp}
  \end{center}
\end{figure}

\end{document}